\documentclass[fleqn,usenatbib]{mnras} 
\usepackage{newtxtext,newtxmath}
\usepackage[T1]{fontenc}
\usepackage{ae,aecompl}
\usepackage{graphicx}	
\usepackage{amsmath}	
\usepackage{amssymb}	
\usepackage[dvipsnames]{xcolor} 

\newcommand{\ronerev}{\color{black}}
\newcommand{\cdm}{\color{black}}

\newcommand{\rhat}{\mathbf{\hat{r}}}
\newcommand{\rvec}{\mathbf{r}}
\newcommand{\rcm}{\mathbf{r}_{\rm cm}}
\newcommand{\Force}{{\mathbf F}}
\newcommand{\Forcerad}{{\mathbf F}_{\rm rad}}
\title[{ Capture and Escape }]{{ Capture and Escape: The Dependence of Radiation Forces on Clumping in Dusty Envelopes }}
\author[P. H. Jumper \& C. D. Matzner]{
Peter H. Jumper,$^{1}$\thanks{E-mail: jumper@astro.utoronto.ca (PHJ)}
C. D. Matzner,$^{1}$ 
\\
$^{1}$Department of Astronomy and Astrophysics, University of Toronto, 50 St. George Street, Toronto, Ontario, M5S 3H4, Canada \\ \rightline{Submitted to MNRAS, 13 June 2018}
}
\date{Accepted XXX. Received YYY; in original form ZZZ}
\pubyear{2018}
\begin{document}
\label{firstpage}
\pagerange{\pageref{firstpage}--\pageref{lastpage}}
\maketitle


\begin{abstract}{  Dust barriers effectively capture the photon momentum of a central light source, but low-density channels, along with re-emission at longer wavelengths, enhance its escape.  We use Monte Carlo simulations to study the effects of inhomogeneity on radiation forces imparted to a dust envelope around a central star.  We survey the strength and scale of an inhomogeneous perturbation field, as well as the optical depth of its spherical reference state. {\cdm We run at} moderate numerical resolution, relying on our previous resolution study for calibration of the associated error.  {\cdm We find that} inhomogeneities matter most when their scale exceeds the characteristic mean free path.   As expected, they tend to reduce the net radiation force and extend its range; however, there is significant variance among realizations.    {\ronerev Within our models, force integrals correlate with the emergent spectral energy distribution, given a specified set of dust properties.} 
{\ronerev  A critical issue is the choice of integral measures of the radiation force:
for} strong deviations from spherical symmetry the relevant measures assess radial forces relative to the cloud centre of mass.  Of these, we find the virial term due to radiation to be the least stochastic of several integral measures in the presence of inhomogeneities.  

}\end{abstract}


\section{Introduction} \label{introduction}

Radiation pressure forces are suggested to play a key role in several contexts where massive stars interact with interstellar matter. The capture of photon momentum by dust grains is responsible for superwind mass loss from asymptotic giant branch stars \citep{1991ApJ...375L..53B}, poses a formidable barrier to massive star formation \citep{wolfire87}, and may be influential in the removal of matter during massive star cluster formation and the disruption of molecular clouds  \citep{2010ApJ...710L.142F,2010ApJ...709..191M}. The effect of radiation forces on H II regions \citep{2009ApJ...703.1352K,draine11,2012ApJ...757..108Y} has been  cited as a crucial precondition for the successful deposition of supernova energy  on galactic scales \citep{hopkinsetal14}, and radiation has been directly implicated in disk support \citep{2005ApJ...630..167T,2009ASPC..408..128T} and galactic winds \citep{2012MNRAS.424.1170Z, 2012MNRAS.425..605F} driven by starbursts or AGN.  At the same time, there are counterexamples in which radiation forces are found to be relatively weak, especially in star cluster formation and molecular cloud disruption \citep{daleetal05,2011ApJ...738...34P,matznerjumper15}.

We wish to delineate the influence of radiation pressure forces, but we must be mindful of several complications.  First, the opacity of dust grains is strongly wavelength dependent, allowing luminosity to escape a region more easily once it is absorbed and re-radiated by grains.   Second, conclusions can depend on the choice of numerical method, as studies of super-Eddington galaxy disks \citep{2013MNRAS.434.2329K,2014ApJ...796..107D} and radiation-dominated accretion disks \citep{2009ApJ...691...16H,2013ApJ...778...65J} demonstrate. 
Lastly, most dusty environments are anisotropic or inhomogeneous, easing the escape of luminosity through channels of low optical depth; this is a critical feature of massive star formation models \citep{2002ApJ...569..846Y,2005ApJ...618L..33K,2009Sci...323..754K}.

We addressed the first two complications in a previous paper \citep[][hereafter JM17]{2018MNRAS.tmp.1713J}, in which we used Hyperion Monte Carlo simulations \citep{robitaille11} of dust transfer through spherically symmetrical envelopes and surveyed physical and numerical parameters while comparing to a separate method \citep[DUSTY:][]{dustyuser}. 
Monte Carlo radiative transfer solutions are formally exact in the limit of very high resolution and numbers of photon `packets'.  We found that poor resolution of the mean free path introduces a predictable error, but that the radiation forces in various optical depth regimes are understandable in terms of direct, scattered, and re-emitted radiation.  

In this work, we extend these models with the introduction of inhomogeneities that break the spherical symmetry of the problem, and explore their influence on the radiation force and its distribution. 
{ These considerations may be particularly relevant in application to analytical approximations of systems undergoing radiative feedback, where the effect of reprocessed radiation may be described in a ``trapping factor'' \citep{2005ApJ...631..792C, 2009ApJ...703.1352K, 2010ApJ...709..191M, matznerjumper15}.   The models conducted within this paper illustrate how such trapping factors may vary with the properties of the clumping in the surrounding region, which may help inform choices in models.  Furthermore, these models may also provide insight into the variation with clumping properties for both where within the region the trapped force is deposited and a summary of how the internal structure of the region is evolving at a given instant as a consequence of these forces.  These properties will be related to the parameters used in this paper as introduced in \S \ref{Force_Parameters}.   }


\section{Physical Problem and Implementation} \label{Physical_Problem}

\subsection{Parent Dust Envelopes} \label{Parent_Dust_Envelopes}

Our analysis involves modifying a centrally-illuminated, spherically symmetric dust dust envelope, which we call the ``parent envelope'', with a clumpy ``contrast field''.  

For the parent envelopes we adopt the truncated power-law dust profiles we analyzed previously (JM17).  These consist of a dust density satisfying $\rho\propto r^{-1.5}$ from an inner radius $r_{\rm in}$ to an outer radius of $4\,r_{\rm in}$.  Dust opacity follows the \citet{2003ARA&A..41..241D} model of carbonaceous silicates, although we simplify the scattering processes to be isotropic.  
The governing parameters for radiation transfer through the parent envelope are: 
\begin{itemize} 
\item[-] $T_*$, the colour temperature of the central light source, set to 5772\,K;
\item[-] The density power law and ratio of outer to inner radii, set to $-1.5$ and 4, respectively; 
\item[-] the innermost dust temperature $T_{\rm in}$, set to 1500\,K in this study;  and 
\item[-] the optical depth to starlight, $\tau_*$, for which we consider values of $10^{-1}$ to $10^{1.5}$ in half-decade increments.   
\end{itemize}
These parameters suffice to determine the flux at  $r_{\rm in}$, as listed in in  Table \ref{tab:parents}.  The 
inner radius $r_{\rm in}$ itself depends on the central luminosity $L_*$, which we arbitrarily set to  $L_\odot$.  The inner dust density is determined by $r_{\rm in}$ and $\tau_*$. 




\begin{table}
\begin{centering}
\caption{Parent Envelopes with {$ T_{\rm in} = 1500 \ {\rm K}$} }
\label{tab:parents}
\begin{tabular}{|c |c |}
\hline {$\log_{10} \tau_*$} & {Flux at $r_{\rm in}$}\\ &  ($10^6$\,erg\,cm$^{-2}$\,s$^{-1}$) \\
\hline
${-1.0}$ & 240 \\
${-0.5}$ & 216 \\
${0.0}$ & 171 \\
${0.5}$ & 125 \\
${1.0}$ &96.8\\
${1.5}$ & 76.6\\
${2.0}$ & 55.6 \\
\end{tabular}
\end{centering}
\end{table}



\subsection{Contrast Fields}  \label{perturb_setup}


We create inhomogeneous dust envelopes by multiplying the parent envelope by a log-Gaussian contrast field with a definite power law spectral { index}. { Our choice of a log-Gaussian (lognormal) field is motivated from the general observation that star-forming environments tend to be turbulent and that the results of numerous numerical studies have shown that such turbulent environments tend to give rise to a log-Gaussian probability distribution function in density (\citealt{2002ApJ...576..870P} and references therein).} {\cdm While we do not develop inhomogeneities self-consistently from the dynamics of turbulence, our approach maintains a level of realism while allowing us to study the influence of a small set of parameters describing the dust density.} 


{ The} procedure { for generating the contrast fields} is  detailed in Appendix \ref{S:ContrastFields}.  It starts with one of six definite random seeds to specify realizations named ``Case A'' through ``Case F''; uses this to generate amplitudes and phases for a Gaussian random field; specifies mode amplitudes according to a power law spectrum ($\propto k^{\beta}$ for $-2\leq\beta\leq-1$ in increments of 0.25).  The field is then multiplied by an amplitude factor ($0\leq q\leq 1$ in steps of 0.2).  This yields a trial contrast field which is  multiplied by the parent envelope density.  Finally,  $\rho(\rvec)$ and its contrast field $\cal P$ are obtained by renormalizing the result to preserve the mass and mean optical depth of the parent envelope.

Models are parameterized by $\tau_*$, the random seed, the spectral slope $\beta$, and the strength of the contrast field.  Rather than use our input parameter $q$, we prefer the `clumping factor' $\langle {\cal P}^2 \rangle/ \langle{\cal P}\rangle^2$ as a measure of its strength. 

A feature of our approach is that the random seed sets the initial phases of the modes that enter $\cal P$, while the spectral slope $\beta$ determines their relative amplitudes.  Therefore, each case (A through F) generates smooth distributions of any force parameter as $q$ or $\beta$ is changed continuously. 

\subsection{Numerical Implementation}\label{Numerics}

We initialize the parent envelopes in the Monte Carlo radiative transfer code Hyperion \citep{robitaille11} at fixed  128$^3$ resolution, using a total of $6.4 \times 10^7$ photon packets, for an average of $\approx 30.5$ photon packets per cell.  As described in JM17, we modified Hyperion to directly calculate the specific radiation force (radiation pressure force per unit mass).  Following from the method of \citet{1999A&A...344..282L}, each given photon packet is assigned a target optical depth to propagate to before an interaction may occur to alter its path, either by scattering or by absorption and re-emission.  During this process, two forms of energy deposition occur: continuous absorption during the propagation along the entire path to the point of interaction, and a localized transfer of momentum at this point based on the difference in energy and velocity (including direction) between the incident photon and the scattered or re-emitted photon that emerges.  The resulting transfers were calculated in terms of the radiation's ``specific force'',  or the force exerted per unit mass (an acceleration).  Integrating the specific force with the properties of the envelope summarizes the force-capturing in the envelope in terms of a group of parameters, as discussed in \S \ref{Force_Parameters}.  


{ Our analysis from JM17 allows us to calibrate the degree to which results quoted here will be affected by the finite resolution of our runs.  Comparing Hyperion runs of varying Cartesian resolution against converged spherical solutions from DUSTY \citep{dustyuser}, we found there that the force will be underestimated at $128^3$ resolution, but by less than 1\% for $\tau_*<5.5$, increasing to a few percent for $\tau_*\sim 10^2$.  These errors are partly due to an under-representation of the inner dust temperature in cases where the starlight mean free path is unresolved.  We do not know how the error depends on inhomogeneity of the dust distribution, but we have no reason to think it should be sensitive.   The force error is quite small compared to the change in force due to clumping, as we shall see in  \S \ref{scaling}.  }

\subsection{Force Parameters} \label{Force_Parameters}

In JM17 we introduced three integral quantities to measures forces:  $\Phi$, the ratio of the total applied outward radiation force $\Forcerad$ to the photon force of the stellar luminosity; $\langle r \rangle_F $, the force-averaged radius; and  $\mathcal{R}$, the radiation term in the virial theorem.   These remain useful for inhomogeneous distributions of dust and luminosity, but we must be careful to adopt the correct origin for the radial vector $\rvec$ that appears in their definitions.  In Appendix \ref{virial_analysis} we demonstrate that the centre of mass (CM) is the correct origin for the definition of $\cal R$, and therefore we define 
\begin{equation} \label{phi_r0}
\Phi = \frac{\int \hat\rvec \cdot d\Forcerad }{L/c},
\end{equation}
\begin{equation}
\langle r \rangle_F = \frac{\int \rvec \cdot d\Forcerad}{\int \hat\rvec \cdot d\Forcerad},
\end{equation}
and 
\begin{equation}
{\cal{R}} = \int\rvec \cdot d\Forcerad = \Phi\, \langle r \rangle_F \frac{L}{c}, 
\end{equation}
where $\hat \rvec = \rvec/|\rvec|$ is the radial unit vector, and   $\rvec$ is measured relative to the CM, i.e., $\rvec_{\rm cm}=0$.  We identify the CM by assuming a constant dust-to-gas ratio.    We also define dimensionless versions of the latter parameters: $\langle \tilde{r} \rangle_F = \langle r \rangle_F/r_{\rm in}$ and $\tilde{\mathcal{R}} = \mathcal{R}/(r_{\rm in}{L}/{c})$.  

Note that $\Phi$ and $\cal R$ will be positive when the radiation force points away from the CM, i.e. when the cloud is illuminated from within, and will tend to be negative when it is illuminated from without.  Note also that only radial forces enter these integrals.  Tangential forces can affect the motion of the CM (see Appendix \ref{virial_analysis}), and can contribute indirectly to the virial theorem for internal cloud motions by enhancing its internal kinetic energy. 

{ These quantities {\cdm characterize the properties of the radiation force}. The ``trapping factor'' used in previous works is, for an optically-thick envelope, $\Phi-1$.  The force-averaged radius $\langle r \rangle_F$ describes where in the envelope the photon force is deposited.  These combine to give  ${\cal{R}}$, the {\cdm net dynamical inflence of radiation as measured by the virial theorem}; see Appendix \ref{virial_analysis}.       }


\section{ Results and Discussion   }   \label{Results_Section}

In our parameter survey, we vary the properties of the dust envelopes for six random seeds, seven optical depths ($\tau_* = 10^{-1}$ to $10^{2.0}$), five spectral indices $\beta$, and six amplitude parameters $q$. This amounts to 1260 combinations, although as all $q = 0$ combinations reduce to the parent envelope, there are only 1057 unique physical problems. 
 


In \S \ref{visualized} we provide examples of envelopes within our parameter space as well as the force, specific force, and temperature within each model.  In \S \ref{scaling} we characterize the scaling of the force-capturing parameters with envelope optical depths and anisotropies.   We connect the force parameters to the emergent spectral energy distribution in \S \ref{SS:SEDs}.

\subsection{Envelope Visualization: Anisotropy Variation and Radiative Capture} \label{visualized}
  
\begin{figure}
\centering
\includegraphics[width=0.50\textwidth]{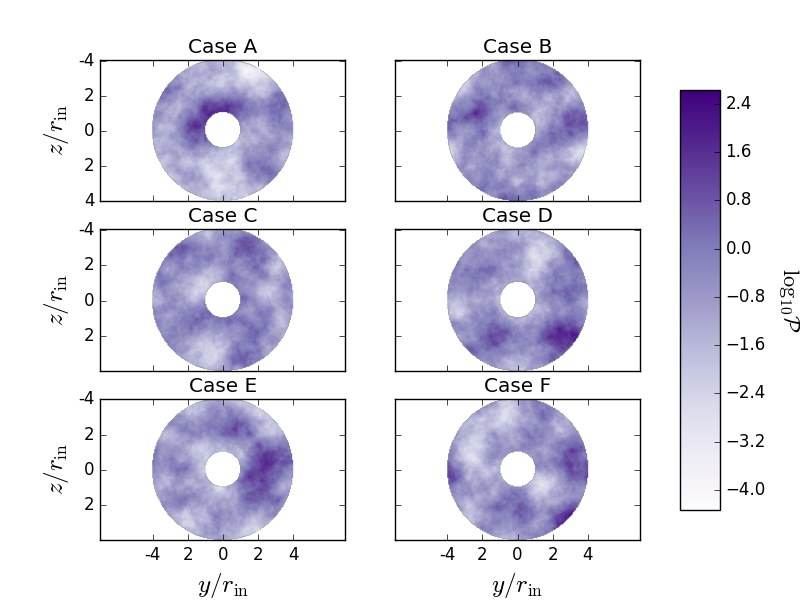} 
\caption{ Slices of the $\log_{10}$ of the contrast field $\mathcal{P} $ through the center of the dust envelope in the yz-plane for the realizations of the six random seeds, labeled Cases A-F, for constant values of $\tau_* = 10.0$, $\beta = -2.00$, and $q = 1.0$.}\label{seed_cases}
\end{figure}

\begin{figure}
\centering
\includegraphics[width=0.50\textwidth]{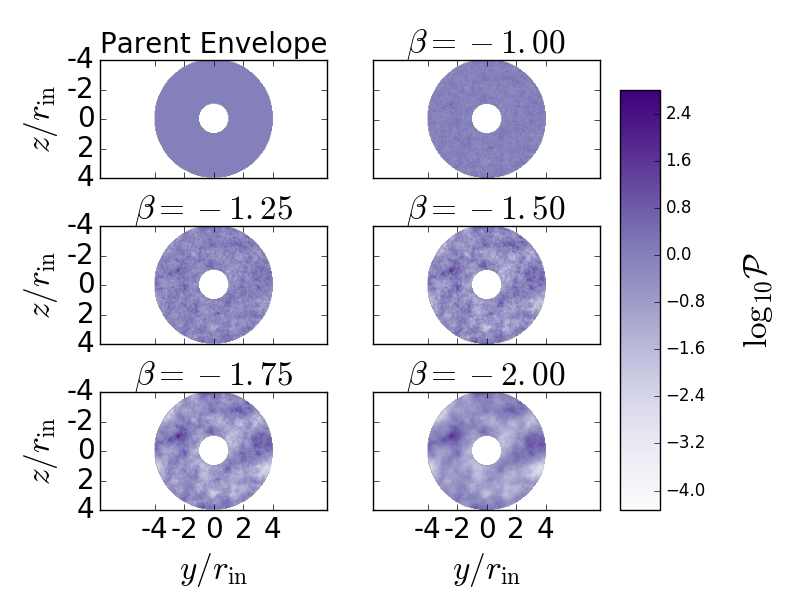}
\caption{ Slices of the $\log_{10}$ of the contrast field $\mathcal{P}  $ through the center of the dust envelope in the yz-plane for the random seed used in Case B of Figure \ref{seed_cases}, holding $\tau_* = 10.0$ and $q = 1.0$ constant and varying $\beta$. Additionally, the unperturbed parent envelope (with $\mathcal{P} =1$ everywhere) is included in the upper-left panel of the figure for reference.  As $\beta$ grows more negative, the spatial extent of the clumps and pores grows as well. }\label{beta_variation}
\end{figure}

\begin{figure}
\centering
\includegraphics[width=0.50\textwidth]{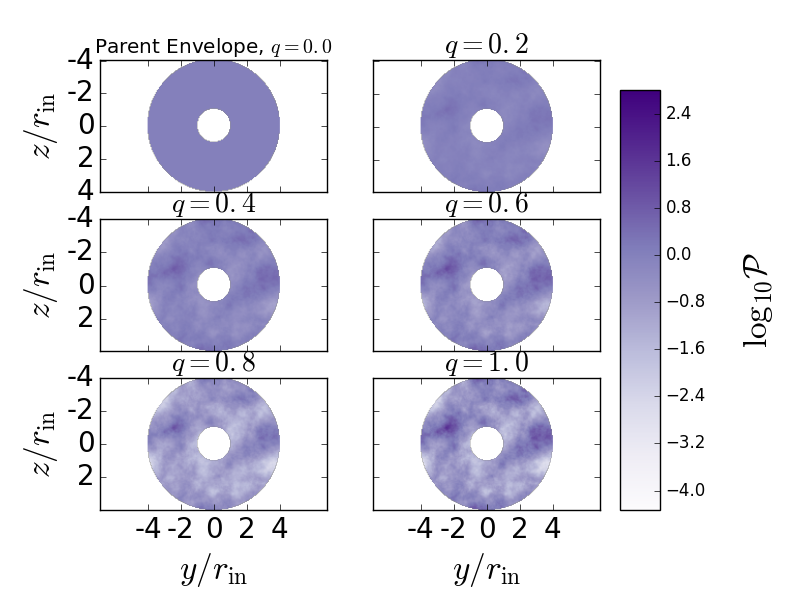}
\caption{Slices of the $\log_{10}$ of the contrast field $\mathcal{P} $ through the center of the dust envelope in the yz-plane for the random seed used in Case B of Figure \ref{seed_cases}, holding $\tau_* = 10.0$ and $\beta = -2.00$ constant and varying $q$. The unperturbed parent envelope (with $\mathcal{P} =1$ everywhere) coincides with the $q =0$ case.  As $q$ grows, the envelopes' clumps become denser and pores become rarer, both while preserving their positions.} \label{psi_variation}
\end{figure}

We illustrate the patterns generated by Cases A--F in Figure \ref{seed_cases}, showing each case for constant values of $\tau_* = 10$, $\beta = -2.0$, and $q = 1.0$.   
We then take one of these cases, Case B, and demonstrate how the contrast factor changes as the other parameters are varied:  $\beta$ and $q$ in Figures \ref{beta_variation} and \ref{psi_variation} respectively. The former illustrates the variation of the spatial extent of the clump structure with $\beta$, with the clump sizes growing as $\beta$ becomes more negative.  The latter varies the amplitude parameter $q$, making the clumps and pores more (larger $q$) or less (smaller $q$) intense while preserving their locations. In the limiting case of $q=0$, $\mathcal{P} = 1$ at all points and the parent envelope is obtained.

\begin{figure}
\centering
\includegraphics[width=0.50\textwidth]{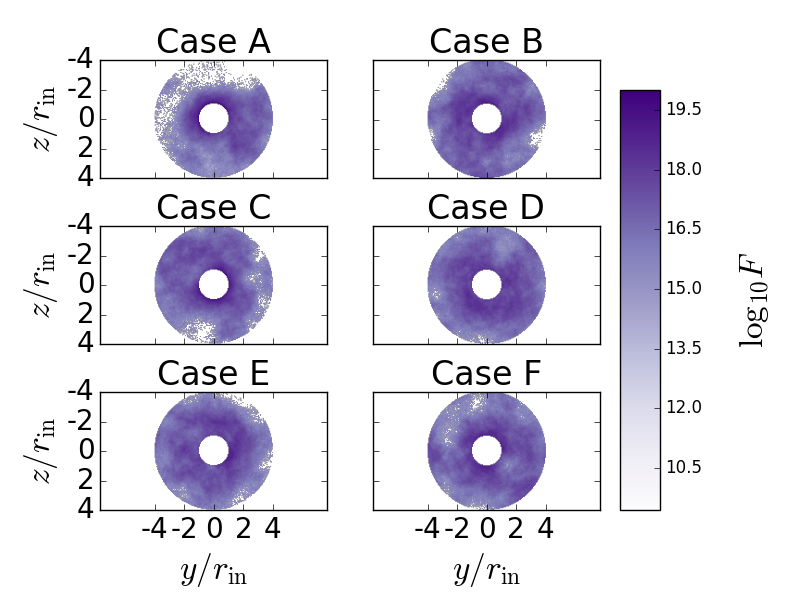}
\caption{ Slices of the $\log_{10}$ of the  force exerted on the dust in units of ${\rm g \ cm \ s^{-2}} $, ${\rm F}$, through the center of the dust envelope in the yz-plane for the realizations of the six random seeds, Cases A-F, as introduced in Figure \ref{seed_cases} holding $\tau_* = 10.0$, $\beta = -2.00$, and $q = 1.0$ constant.  Some regions were heavily shielded by clumps and received no photons.}\label{force_slices}
\end{figure}

\begin{figure}
\centering
\includegraphics[width=0.50\textwidth]{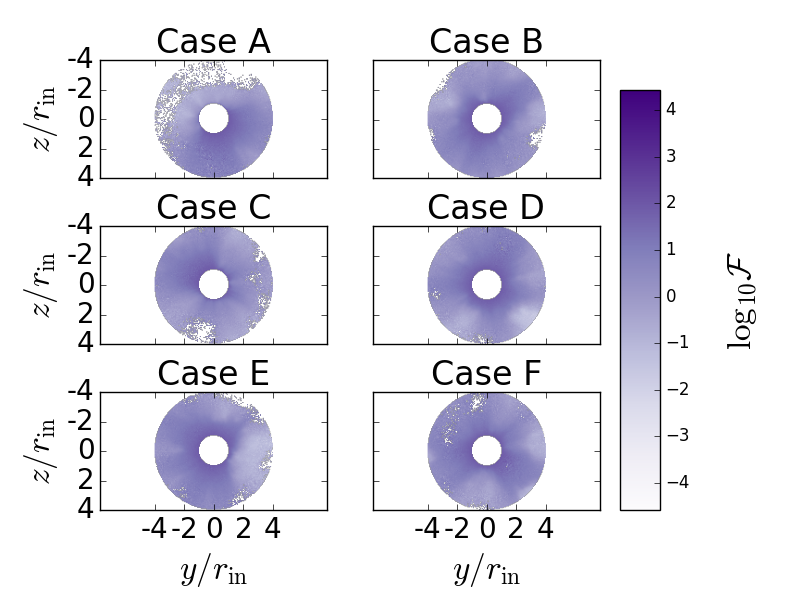}
\caption{  Slices of the $\log_{10}$ of the specific force (acceleration) exerted on the dust in units of ${\rm cm \ s^{-2}} $, $\mathcal{F}$, through the center of the dust envelope in the yz-plane for the realizations of the six random seeds , Cases A-F, as introduced in Figure \ref{seed_cases} holding $\tau_* = 10.0$, $\beta = -2.00$, and $q = 1.0$ constant.  Some regions were heavily shielded by clumps and received no photons. }\label{specific_force_slices}
\end{figure}

\begin{figure}
\centering
\includegraphics[width=0.50\textwidth]{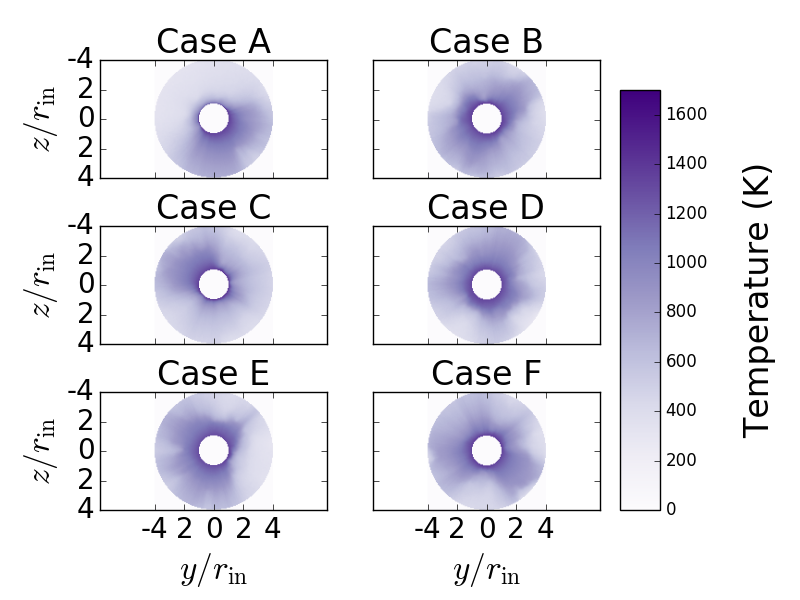}
\caption{ Slices of the dust temperature (K) through the center of the dust envelope in the yz-plane for the realizations of the six random seeds , Cases A-F, as introduced in Figure \ref{seed_cases} holding $\tau_* = 10.0$, $\beta = -2.00$, and $q = 1.0$ constant. The physical dimensions of the dust envelope were set to establish an inner wall temperature of 1500 K for the parent envelopes, as calculated by DUSTY, and the temperature determination routine has a floor temperature of 0.1 K.    } \label{temperature_slices} 
\end{figure}

We conducted a Monte Carlo model, using Hyperion, to solve the radiative transfer for each dust envelope in our survey, obtaining the specific forces and temperatures in each cell.  These forces were then used to calculate the force-capturing parameters ($\Phi$, $\langle r\rangle_F$, and $\mathcal{R}$) and their dimensionless forms.   
We also illustrate the force, specific force, and temperatures, shown in Figures \ref{force_slices},  \ref{specific_force_slices},  \ref{temperature_slices}, calculated for slices of constant $\tau_* = 10$, $\beta = -2.0$, and $q = 1.0$, varied over the different seeds, to provide an example of their realizations. Relating the forces seen in these figures to the clumping $\mathcal{P}$ shown in Figure \ref{seed_cases}, we observe that the total force captured per cell strongly correlates with highly overdense or `clumped' (large $\mathcal{P}$) regions (Figure \ref{force_slices}), while the specific force (acceleration) in the cells tracks with strongly underdense or `porous' ($\mathcal{P} << 1$) regions instead (Figure \ref{specific_force_slices}). Although our models are static, this suggests that in a dynamical model the radiation forces would act to intensify the existing density contrasts, as the material in lower-density porous regions will accelerated more strongly and preferentially driven away.  In addition, we find that the temperature, calculated as a function of the dust's specific energy absorption rate, also is higher in the porous region and lower in both highly clumpy regions and the shadowed regions exterior to them (Figure \ref{temperature_slices}).

We note that Figures \ref{force_slices} and \ref{specific_force_slices} indicate that some regions, particularly where there is heavy shadowing from prominent clumps to the interior, received no photon packets (and thus no forces).  While running the model with a larger number of photon packets may have allowed packets to reach these regions, the fact that they received none of the $6.4 \times 10^7$ photon packets used in the model (representing an average of about 30.5 packets/cell for the $128^3$ box) suggests that the contributions of these regions to the overall solution is negligible.  This absence is not seen in Figure \ref{temperature_slices}, where the Hyperion temperature function has a floor of $0.1 \ {\rm K}$.     

\subsection{Influence of Inhomogeneities  on the Radiation Force} \label{scaling}

We now present our results on the influence of inhomogeneities on the capture of radiative forces, as summarized in the parameters' dimensionless forms: $\Phi$, $\langle \tilde{r} \rangle_F$, and $\tilde{\mathcal{R}}$ .  

We begin with Figures \ref{two_panel_phi}, \ref{two_panel_rf}, \ref{two_panel_y}, which plot the variation of our parameters against a ``clumping factor'', $\langle \mathcal{P}^2 \rangle/\langle \mathcal{P} \rangle^2$ (right panels), and relate these to the variation with $\tau_*$ in the parent envelopes (left panels).  The curves on the left panels are computed with another code, DUSTY, as described in JM17.  Our clumping factor is analogous to the definition by \citet{2018MNRAS.475..814O}. 

{ Insofar as converged DUSTY results align with the high-resolution limit of Hyperion runs (a result from JM17), the effect of finite resolution can be seen in these figures as a slight vertical offset between the dashed lines on the left panel and the dots on the right panel.  This offset is barely visible, and clearly quite small. }


In these figures, all curves sharing the same seed are plotted with the same colour, and all curves plotted display six values at a given optical depth for the parameter in question, holding $\beta$ and seed values constant and allowing the amplitude parameter $q$ to vary, which in turn also varies $\langle \mathcal{P}^2 \rangle/\langle \mathcal{P} \rangle^2$.   
We see that in many cases the scaling can be approximated as a power law in $\langle \mathcal{P}^2 \rangle/\langle \mathcal{P} \rangle^2$. We discuss such power-law fits
later in the paper.  

We observe that in the majority of cases, $\Phi$ and  $\tilde{\mathcal{R}}$ diminish as a function of the contrast clumping factor, while  $\langle \tilde{r}\rangle_F$ increases.  The results for $\Phi$ and  $\langle \tilde{r}\rangle_F$ are as expected, as the concentration of matter into clumps leaves underdense porous regions in the spaces between them, which allow photons to easily leak through to escape, thus bypassing the locally optically thicker regions established by the clumps. Thus, less force tends to be captured overall, and the capturing which does occur tends towards further out radii.  Regarding $\mathcal{R}$, in the limiting case of a spherically symmetric envelope, $\mathcal{R} = \Phi  \langle r \rangle_F {L}/{c}$, so whether $\mathcal{R}$ increases or decreases with the introduction of anisotropies depends on whether $\Phi$ or $\langle r \rangle_F$ scales more strongly and dominates the calculations. A question posed in JM17 is now answered: we see in Figure {\ref{two_panel_y} that $\tilde{\mathcal{R}}$ diminishes with the introduction of clumping.   

However, we also observe that in the cases of the realizations from certain seeds, the scaling behaviors behave differently from the manner described above.  This is seen particularly prominently in Figures \ref{two_panel_phi} and \ref{two_panel_rf} ($\Phi$ and  $\langle \tilde{r}\rangle_F$), especially in Case A (red curves) to a lesser extent in Case C (gold curves).  Figure \ref{seed_cases} helps us understand why this anomalous behavior occurred.  
We see that in both cases the clumps are positioned such that much of the inner cavity wall, where the photons first arrive at the dust envelope, abuts these clumps, although much more prominently in Case A.  
The local increase in optical depth leads to a net increase in force.  
Although in both cases there are lower-density regions on other sides of the cavity wall, it appears in both cases the particular clump geometry and position has produced a net enhancement in the force capturing, in contrast to the reduction found in the other cases.  Returning to Figure \ref{two_panel_phi}, we also note that this effect is most prominent when the parent envelope is optically thin and diminishes as the parent envelope becomes optically thick.  This is most likely attributable to high contrast clumps creating locally optically thick regions in the anisotropic envelopes that readily capture photons that would have easily escaped from the thin parent envelopes.   

Observing these figures, we also find that $\mathcal{R}$ is less sensitive to dependence on the particular realization of the generating seed that the other parameters, as seen in the much narrower dispersion of its curves.  This outcome makes sense, 
as 
$\mathcal{R} = \Phi \langle r \rangle_F L/c$.   As has been established, $\Phi$ tends to decrease as anisotropies are introduced, while $\langle r \rangle_F$ behaves in the opposite manner and tends to increase.  Any variations in a particular realization of the effects these anisotropies have on the radiative capturing from the typical result for a given set of properties will then tend to affect these parameters in the opposite directions, so when we consider that they may be considered as factors of $\mathcal{R}$, there is a greater tendency towards these variations largely canceling.  Thus the reduced dispersion in solutions for $\mathcal{R}$ is reasonable. 


\begin{figure}
\centering
\includegraphics[width=0.50\textwidth]{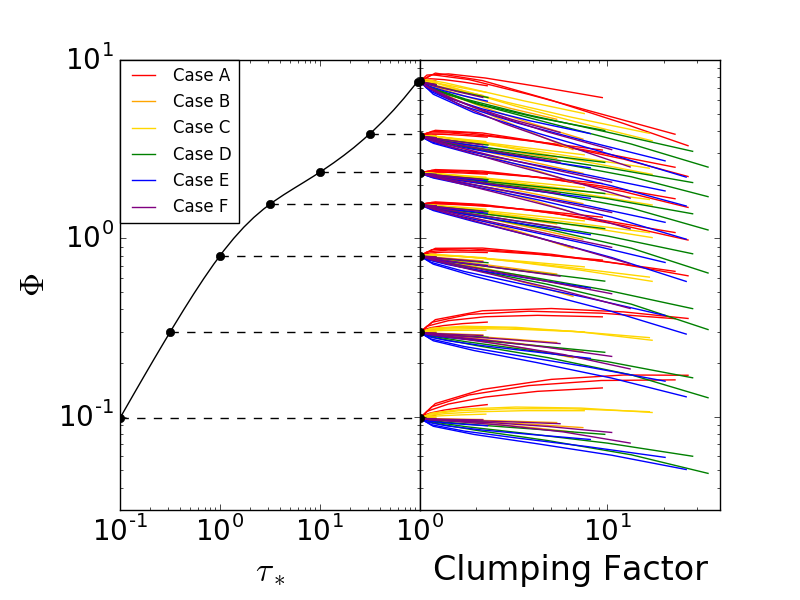}
\caption{ {\em Left:} DUSTY models for $\Phi$ for a collection of envelope optical depths $\tau_*$ from JM17 (solid line) and for this parameter survey (dots). {\em Right:} The variation of $\Phi$ with the clumping factor, $\langle \mathcal{P}^2 \rangle/\langle \mathcal{P} \rangle^2 $,  for all models in the parameter space.  All models of the same random seed case are plotted with the same colour. { In this and the two following figures, any error due to finite resolution is visible as a vertical offset between dots in the left and right panels.}} \label{two_panel_phi} 
\end{figure}

\begin{figure}
\centering
\includegraphics[width=0.50\textwidth]{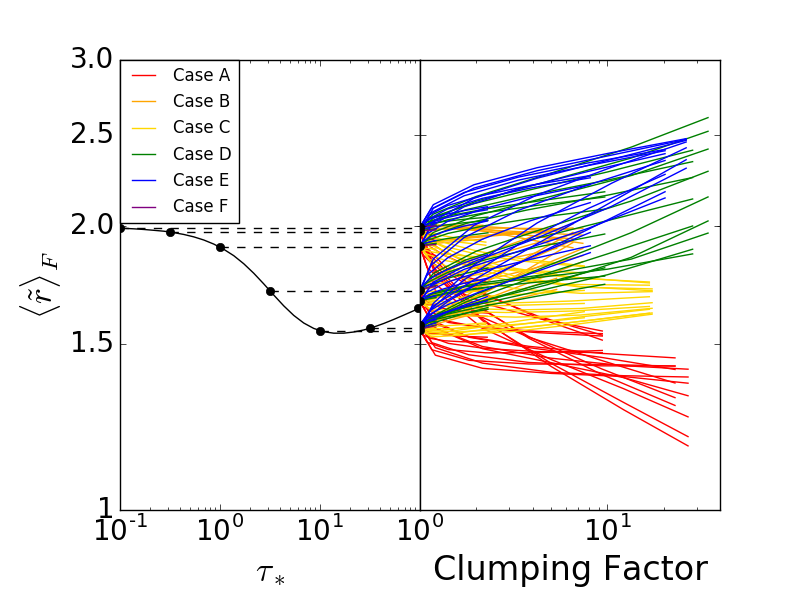}
\caption{ {\em Left:} DUSTY models for  $\langle \tilde{r} \rangle_F/$ for a collection of envelope optical depths $\tau_*$ from JM17 (solid line) and for this parameter survey (dots). {\em Right:} The variation of $\langle \tilde{r} \rangle_F/$ with the clumping factor, $\langle \mathcal{P}^2 \rangle/\langle \mathcal{P} \rangle^2 $, for all models in the parameter space.  All models of the same random seed case are plotted with the same colour.  } \label{two_panel_rf}
\end{figure}

\begin{figure}
\centering
\includegraphics[width=0.50\textwidth]{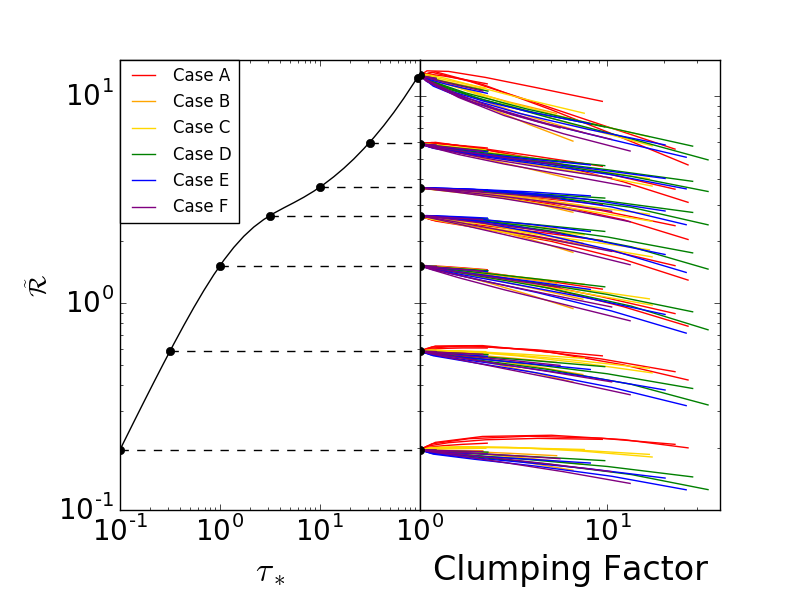}
\caption{{\em Left:} DUSTY models for  $\tilde{\mathcal{R}}$  for a collection of envelope optical depths $\tau_*$ from JM17 (solid line) and for this parameter survey (dots). {\em Right:} The variation of  $\tilde{\mathcal{R}}$ with the clumping factor, $\langle \mathcal{P}^2 \rangle/\langle \mathcal{P} \rangle^2 $, for all models in the parameter space.  All models of the same random seed case are plotted with the same colour.   } \label{two_panel_y}
\end{figure}

\begin{figure}
\centering
\includegraphics[width=0.50\textwidth]{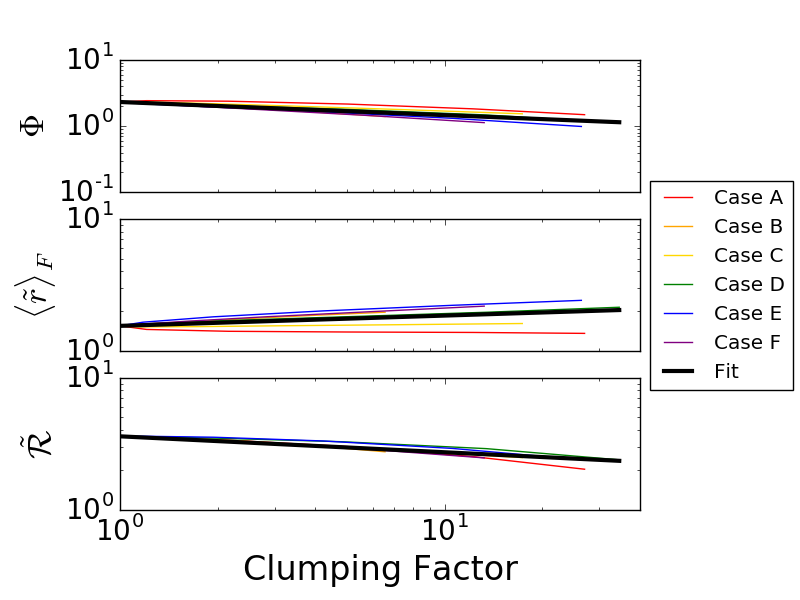}
\caption{For $\tau_* = 10.0$ and $\beta = -2.00$, an examination of the dependence on the three force-capturing parameters on the clumping factor.  The black line represents an overall power-law fit to the data, derived from the average index of the power-law fits for seed Cases A-F.  }
\label{group_32_fit}
\end{figure}

We now consider power-law fits of the form $\Phi = \Phi_0(\tau_*) \left(\langle \mathcal{P}^2 \rangle/\langle \mathcal{P} \rangle^2 \right)^{b_\Phi}$, and likewise for $\langle r \rangle_F$ and $\cal R$.  We average these power-law indices ($\beta_\Phi$, $\beta_r$, and $\beta_{\cal R}$, respectively) over random seeds to explore trends with $\beta$ and $\tau_*$.  
We demonstrate the resulting fitting in Figure \ref{group_32_fit} for the $\tau_* = 10.0$ and $\beta = -2.00$ for each force-capturing parameter. 

We conducted this procedure for all 35 $\tau_*$-$\beta$ combinations, each of which carried the information for 36 parameter models (varying over 6 seeds and 6 amplitude factors $q$).  We present these results in the form of contour plots for the power-law fit indexes, shown in Figures \ref{phi_contour}, \ref{rf_contour}, and \ref{y_contour}.  

We observe that the power-law fit to the scaling of $\Phi$ with clumping becomes more negative with increasingly negative $\beta$ (and thus a larger characteristic size scale of the clumps). This result is reasonable; the larger clumps come alongside larger channels between the clumps, which facilitates the easier escape of photons between these channels.  The opposite is true for small scale pores with less negative $\beta$ values.  

In the lower-left corner of the figure, where $\beta$ and $\tau_*$ have their smallest magnitudes, we also see that there is a regime where the power-law index of fit is a very weakly positive value, rather than a negative value.  As discussed previously, Case A and to a lesser extent Case C exhibited a similar behavior in the low $\tau_*$ regime, enhancing the force capturing rather than diminishing it.   However, the remaining seed realizations, Cases B, D, E, and F, behaved in the opposite manner, reducing the force capture as the clumping increased. These opposing sets of values nearly cancel in the averaging, so there is little net effect of clumping in this regime.  

We also observe that the index of fit for $\Phi$ also becomes more negative (the force diminishes more strongly with increasing clumping) as we move towards larger optical depths in the regime from optically thin envelopes to envelopes of several optical depths. This also makes intuitive sense, as clumping should have no effect in an optically thin envelope, while channeling is important in an optically thick one. 

However, 
we observe a deviation from this trend, with a local minimum in the strength of the scaling ($b_{\phi}$ becomes less negative)
around $\tau_* = 10.0$. 
This is probably an effect of the optical depth, as the infrared opacity, being  an order of magnitude lower than the optical opacity, is about unity in this regime.  However we have not ruled out the possibility that resolution effects also play a role; these depend on the resolution of the mean free path, as we discussed in JM17. 
The anomalous behavior is observed in all three of our force parameters (Figures \ref{phi_contour}-\ref{y_contour}).  

Returning to the remaining figures, Figure \ref{rf_contour} shows that the scaling of the dimensionless force-averaged radius, $\langle \tilde{r} \rangle_F $ also intensifies with increasing envelope optical depth over the same regimes as found for $\Phi$; as before, the presence of channels to bypass otherwise thick regimes means that their effect has increased over thin areas.  However, in contrast to our above example, we find that the scaling depends much more weakly on the size of the clumps as characterized by $\beta$.   
In Figure \ref{y_contour},  we see that $\tilde{\mathcal{R}}$ behaves in a manner very similar to that already described for $\Phi$.

\begin{figure}
\centering
\includegraphics[width=0.50\textwidth]{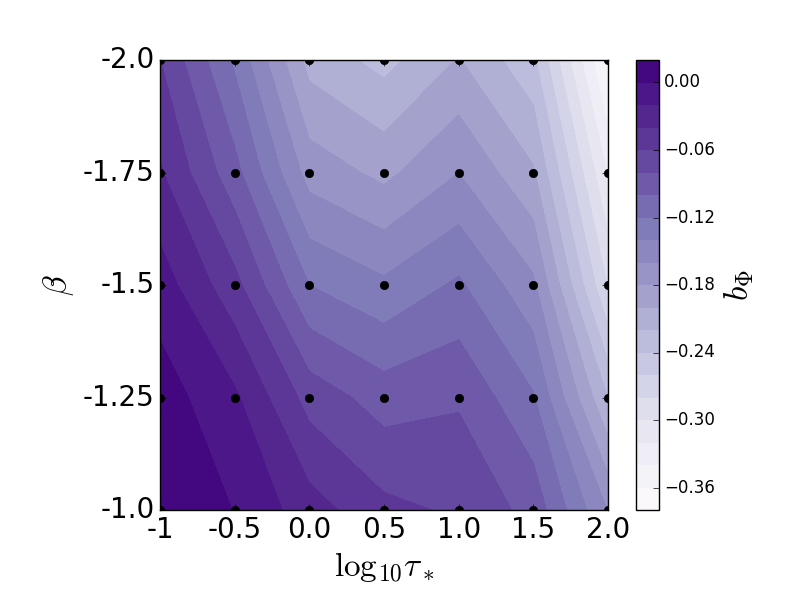}
\caption{ Contour plot of the average power-law exponent, $b_{\Phi}$ , for the scaling of $\Phi$ with the clumping factor, $\langle \mathcal{P}^2 \rangle/\langle \mathcal{P} \rangle^2 $,varying with both $\log_{10} \tau_*$ and $\beta$.  We indicate the parameter values of our data points with black dots and interpolate the contour lines from these data points.   } \label{phi_contour}
\end{figure}

\begin{figure}
\centering
\includegraphics[width=0.50\textwidth]{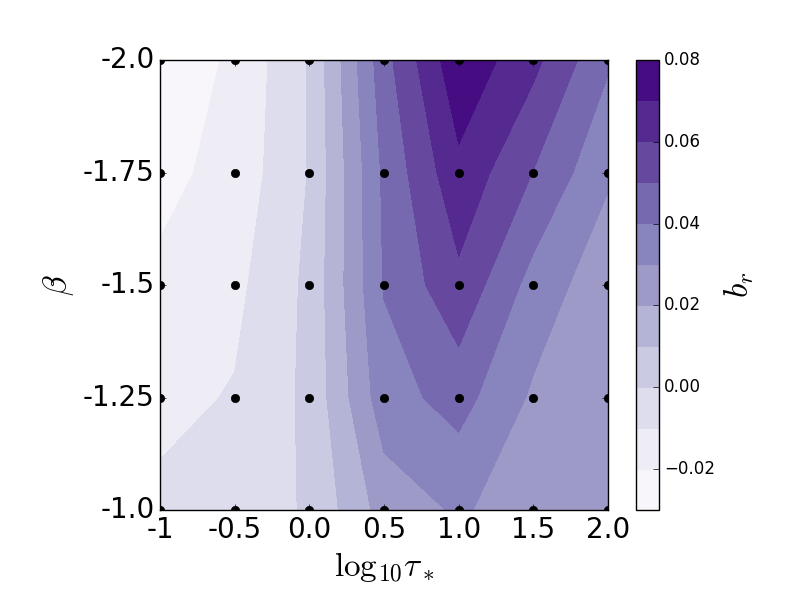}
\caption{  Contour plot of the average power-law exponent, $b_r$ ,  for the scaling of $\langle \tilde{r} \rangle_F/$ with the clumping factor, $\langle \mathcal{P}^2 \rangle/\langle \mathcal{P} \rangle^2 $, varying with both $\log_{10} \tau_*$ and $\beta$.  We indicate the parameter values of our data points with black dots and interpolate the contour lines from these data points.     } \label{rf_contour}
\end{figure}

\begin{figure}
\centering
\includegraphics[width=0.50\textwidth]{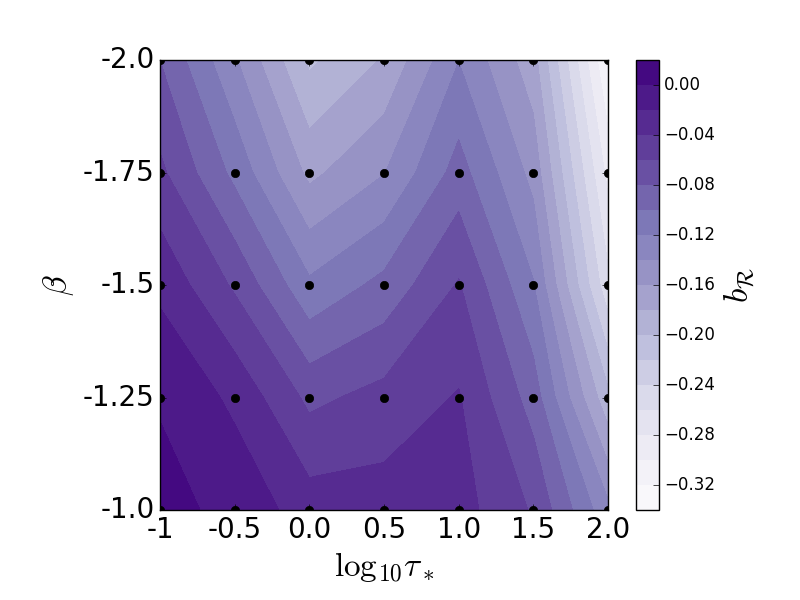}
\caption{  Contour plot of the average power-law exponent, $\beta_\mathcal{R}$ for the scaling of $\tilde{\mathcal{R}}$ with the clumping factor, $\langle \mathcal{P}^2 \rangle/\langle \mathcal{P} \rangle^2 $, varying with both $\log_{10} \tau_*$ and $\beta$.  We indicate the parameter values of our data points with black dots and interpolate the contour lines from these data points.     } \label{y_contour}
\end{figure}

\subsection{SEDs from the Model Envelopes}\label{SS:SEDs}

We now briefly consider the spectral energy distributions (SEDs) that would be observed from our model dust envelopes.  {\cdm We are motivated by the observation that, when inhomogeneities are strong enough to affect the force integrals, they also tend to change the SED by allowing relatively short-wavelength radiation to escape; however, our results cannot be considered diagnostics because we only consider one dust model, and we neglect the possibility of protostellar outflow cavities \citep{WhitneyHartmann93}.  }  All SEDs presented in this subsection may be scaled as a function of the input stellar luminosity and the distance of the observer from the envelopes; we adopt fiducial values of $L = L_{\odot}$ and $d = 10 {\rm pc}$.  For each model in our parameter space, the SED is calculated for 108 wavelengths and six orthogonal viewing angles. 

In Figure \ref{clump_bands}, we examine how the SED varies with the clumping factor $\langle \mathcal{P}^2 \rangle / \langle \mathcal{P} \rangle^2$ for three optical depths ($\log_{10} \tau_* = {-0.5, 0.5, 1.5} $) and for $\beta = {-1, -2} $.   The depths presented here are chosen to represent the three key regimes of the radiative transfer problem: the thin regime, the intermediate regime (thick to direct starlight radiation but thin to reprocessed radiation), and the thick regime. As expected, the peak of the SED is shifted to longer wavelengths from reprocessing as we progress from the thin to thick regimes, and the reprocessing feature around 10 $\mu{\rm m}$ becomes less distinct in this thick regime, where the initially reprocessed radiation itself undergoes further reprocessing. However, as seen in the right side of the figure, in envelopes with large clumping factors, less reprocessing occurs, as some photons are able to escape through the pores.  In turn, more spectral energy is found at these cases closer to the original peak wavelengths, an outcome which intensifies as the clumping factor (and thus also the prominence of the pores between the clumps) increases.  

Additionally, we consider in Figure \ref{sed_figure} how $\tilde{\mathcal{R}}$ varies with the SED slope,  $\log \left( {  \lambda F_\lambda  [10.2 \ \mu{\rm m}] }/{ \lambda F_\lambda [1.02 \ \mu{\rm m}] } \right)  $, to gain insight regarding what an observer may be able to tell about $\tilde{\mathcal{R}}$ from observing the SED.  The particular wavelengths chosen for this slope in our figure coincide with the closest points on the wavelength grid realized by our Hyperion models to $10 \ \mu{\rm m}$ and $1 \ \mu{\rm m}$.  We observe that for optically thin regions, $\tilde{\mathcal{R}}$ varies largely independently of the SED slope; while increasing the clumping factor will tend to decrease $\tilde{\mathcal{R}}$, the SED slope will not vary significantly.  This outcome could be anticipated, as the change in the SED slope between points is driven by reprocessing, which is not a dominant process in optically thin regions.  However, in contrast, as we progress to the optically thick regions, we find a much greater variation of the SED slope with the clumping factor; this accords with our expectations, as well as from our consideration of the SEDs presented in Figure \ref{clump_bands}.

Thus, for optically thick regions, examining the SED slope for the region may offer insight into the intensity of the clumping and the radiative virial parameter within that region if the optical depth of the region { and the dust properties are} known.  However, as a key caveat, since this analysis has been conducted for a particular test model geometry and anisotropic variations away from it, the precise outcomes may vary in more complex environments.

\begin{figure}
\centering
\includegraphics[width=0.50\textwidth]{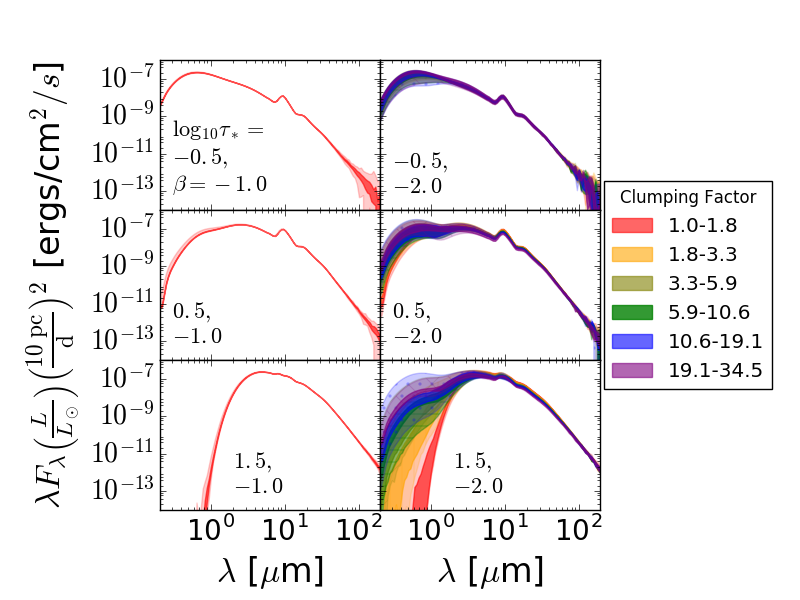}
\caption{For three selected parent envelope optical depths,  $\log_{10} \tau_* = \lbrace{ -0.5, 0.5,1.5 \rbrace}$, and two values $\beta = \lbrace{ -1.0, 2.0 \rbrace}$, $\lambda F_\lambda$ for model SEDS across a range of clumping factors in the fiducial case of a stellar luminosity $L = L_{\odot}$ and a distance from the observer of $10 \ {\rm pc}$.  We organize the data into six bands in the clumping factors, each of equal logarithmic intervals (with a multiplicative factor of $\approx 1.804$ across the band).  We then darkly shade the region between the 25th and 75th percentiles within each band, and lightly shade and hatch the regions outside this interval. The listed band boundaries in the legend have been rounded to one decimal place.    }    
\label{clump_bands}
\end{figure}

\begin{figure}
\centering
\includegraphics[width=0.50\textwidth]{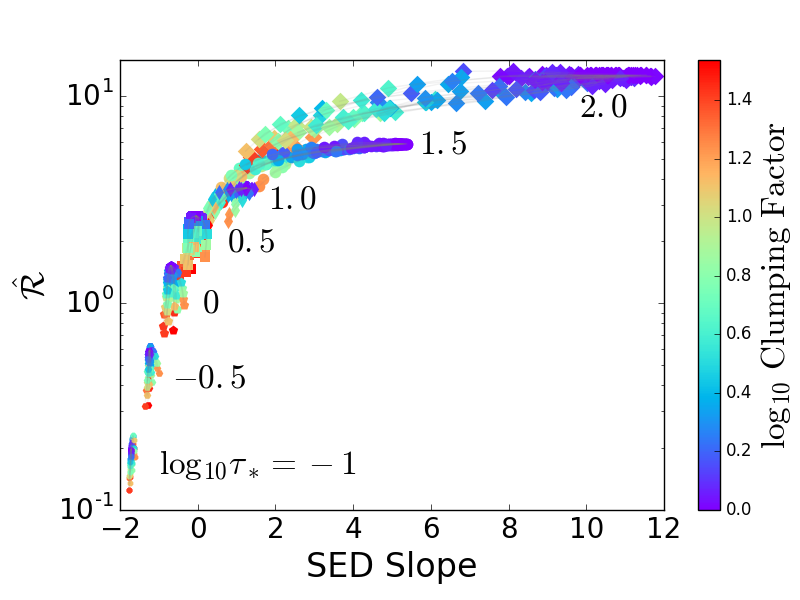}
\caption{Plot of $\tilde{\mathcal{R}}$ vs.\ the SED slope from 1-10\,$\mu$m, 
calculated from the average of the ratio over six viewing angles,  across the parameter space.  We label the regimes for each parent envelope optical depth within our space, and assign each its own data marker.  The colour of the data point shown indicates the logarithm of the their clumping factor ($\langle \mathcal{P}^2 \rangle/\langle \mathcal{P}\rangle^2$). Points with the same $\tau_*$,  $\beta$, and random seed are connected with a thin line. 
}
\label{sed_figure}
\end{figure}

\subsection{Relation to Prior Works} \label{prior_works}

We now take a brief digression to relate our work in this paper to that of prior studies in the field of radiative transfer in clumpy environments. In this paper, we randomly generate a density contrast field spanning over several orders of magnitude and apply this field to an initially power-law density distribution.  This contrasts with the popular ``two-phase'' approximation used in papers such as \citet{1998A&A...340..103W}, \citet{1999ApJ...523..265V}, and \citet{2015A&A...584A.108S}, which in its archetypal form assigns a distinct, single density to the clump regions and another, lower density to the porous regions surrounding them, greatly simplifying the mathematical descriptions of these regions. In addition to the two-phase approximation, some papers further characterize the behavior of radiative transfer in these environments with the mega-grain model, which treats dense clumps like very large dust particles with an associated cross-section for interaction with radiation \citep{1999ApJ...523..265V, 1999astro.ph..5291V}.  Other models utilize continuous stochastic \citep{2003MNRAS.342..453H}, fractal  \citep{1999astro.ph..5291V, 2009PhyA..388.3695W}, or simulation-generated \citep{2018MNRAS.475..814O}  distributions that provide more varied density fields. Our randomly generated density contrast fields are more closely related to this latter set of models.

A key difference our work and that presented in these previous papers is the quantities that are focused upon. Generally speaking, the emphases of the above cited papers has been the determination of an effective optical depth for the modeled regions and/or characterizing the transmission of fluxes or spectral energy distributions through these regions.  In contrast, this paper specifically focuses on the effects of the clumping on the radiation pressure forces captured within the volume of the dust envelope.    
We find that clumping is partly, but not completely, equivalent to a reduction in the overall optical depth. 

\subsection{Conclusions} \label{conclude}

We begin with our key findings.  First, the capture of the radiation force in dust envelopes may be summarized with three integral quantities:  a normalized force, $\Phi$, a force-averaged radius, $\langle r \rangle_F$, and a radiative virial term, $\mathcal{R}$.  Each of these makes reference to the centre of mass, so radiative forces alone do not suffice to determine them; one must also know the mass distribution.  We adopt a constant gas-to-dust ratio when specifying the CM, but other choices are possible.  


Second, as we introduce inhomogeneities into the dust envelope, the parameters $\Phi$, $\langle r \rangle_F$, and $\mathcal{R}$ will vary as a function of both the strength and scale of the perturbations, as well as the base optical depths of the originating symmetric envelopes. We find that these variations can be approximated as power-law behaviors scaling with the degree of introduced clumping, characterized as $\langle \mathcal{P}^2 \rangle/\langle \mathcal{P} \rangle^2$, where $\mathcal{P}$ is the contrast factor between the inhomogeneous and base states in each cell.  Generally, the normalized force $\Phi$ diminishes with increasing clumping, as the porous regions formed between the clumps facilitates the leaking of radiation from the envelope. These porous regions also lead to an increase in the force-averaged radius, as the photons that still are captured tend to leak further before this occurs. The radiative virial term, $\mathcal{R}$ is proportional to $\Phi \langle r \rangle_F$ and tends to vary like $\Phi$ because the variation of $\langle r \rangle_F$ is generally weaker.  However, $\mathcal{R}$ is significantly less sensitive to variations in the clumping caused by realizations of our random contrast field. 

Third, the sensitivity of our force measures to clumping depends on the physical scale of the clumps as well as the overall optical depth. These scalings are demonstrated in contour plots in Figures \ref{phi_contour}, \ref{rf_contour}, and \ref{y_contour}. The scaling tends to be at weakest for all parameters in optically thin envelopes, where most photons would escape regardless of whether or not the clumping was present. In contrast, the scaling tends to increase as we increase the optical depths of the envelopes, as the porous regions introduced alongside the clumping provide bypasses through which the photons may leak. 

However, for all three force-capturing parameters studied, a temporary reversal is found in this trend in the region of $\tau_*\sim 10$ before returning to its previous behavior at even larger optical depths.  This is probably due to the fact infrared dust radiation is marginally optically thick in this regime, although finite-resolution effects could also be contributing.  

The power-law scaling of the force parameters also weakens for small-scale inhomogeneities and strengthens for large-scale inhomogeneities, for $\Phi$ and $\mathcal{R}$, while in contrast $\langle r \rangle_F$ is fairly insensitive to the scale of the clumping for regimes below a few optical depths.  Nonetheless, as the scaling remains relatively weak throughout the entire parameter space, strong clumping is necessary to have a major impact on the capture and escape of radiation in the envelope.  

We briefly consider the variation of $\mathcal{R}$ and the clumping factor with the mid-to-near-infrared SED slope. There is little relation between these in the optically thin regime. In the thick regime the SED slope may offer insight into the degree of clumping and the virial term $\mathcal{R}$  in the envelope,
{\cdm although outflow cavities and uncertainties in the dust model would also affect the emergent SED. }
Insofar as our model represents real clumps, infrared radiation provides some information on the force parameters.  However we have not considered effects, like outflow channels, that may dominate the emergent SED while also affecting the radiative forces. 

In this study, we conducted our radiative transfer for static realizations, and have not considered the dynamical evolution of envelopes.  However, we have observed, as demonstrated in Figures \ref{force_slices} and \ref{specific_force_slices}, that while the force captured per cell is higher in clumpy regions, the specific force (acceleration) exerted upon each cell is larger in lower-density, porous regions.  This observation may become significant in dynamical modeling, where its sets the expectation that the material remaining in the porous space between the clumps will also be the material most strongly accelerated by the radiation forces. In turn, this should contribute to the further evacuation of material from the porous inter-clump regions, either expelling the material or driving it into new clumps. Thus, we may anticipate that the overall degree of clumping in the region should intensify over time, and with it the impact of this on the further capture and escape of photons.


\section*{Acknowledgements}

This work was supported by a {\ronerev Connaught International Scholarship} ({\ronerev Jumper}) and a {\ronerev Natural Sciences and Engineering Research Council of Canada (}NSERC{\ronerev )} Discovery Grant ({\ronerev Matzner}) and enabled in part by support provided by Compute Ontario and Compute Canada for calculations done in the SciNet facility in Toronto.    

\appendix


\section{Contrast Fields}\label{S:ContrastFields}

In this subsection, we will outline our procedure for generating our contrast fields, labeled as $\mathcal{P}(r)$, which we apply multiplicatively to the parent envelopes to introduce inhomogeneities, yielding a new density field, $\rho_{\mathcal{P}}(r) = \rho(r) \mathcal{P}(r)$.  In all cases, we impose the condition that for a parent envelope of mass $M_{\rm initial}$, the new asymmetrical envelope produced from the application of the contrast field conserves the original mass, such that $\int \rho_{\mathcal{P}}(r) dV = M_{\rm initial} $.   

We generate $\mathcal{P}(r)$ as realizations from a set of random seeds, $\mathcal{S}$, in a procedure taking direction from \citet{lewis2002iterative} and an iterative log-normal random density field generator \footnote{\url{http://www.physics.usyd.edu.au/~tepper/software/fractalClouds/doc/html/index.html}, accessed August 24, 2017}. In this paper, we adopt $\mathcal{S} = \lbrace 0, \ 10, \ 20, \ 30, \ 40, \ 50  \rbrace$ for our seeds, which we shall herein refer to Cases A-F respectively, to initialize the Python random number generator. Once so initialized, we generate an $N^3$ cell cube of Gaussian random noise with a mean value of $\mu = 0.0$ and $\sigma = 0.01$; in this paper, $N = 128$.  This random noise provides the basis for the underlying patterns realized in the contrast field $\mathcal{P}(r)$.   

We then take a numerical fast Fourier transform of this Gaussian noise, converting it to a wavenumber space.  We then also generate a spectrum  $\mathcal{E}$ for the wavenumber space which we will multiply against this noise.  The purpose of this spectrum is to impose a characteristic scale to the physical extent of the clumps produced by the noise; by varying the index given to this spectrum, we can produce smaller or larger clump structures.   To this end, we choose a corner of the wavenumber space cube to serve as an origin, then for that corner's octant measure the distance the distance in cell lengths of each cell from that corner as $k$. We then assign  $\mathcal{E} = k^\beta$ for all cells in that octant; for this paper, we consider the cases of $\beta = \lbrace{-1, \ -1.25, \ -1.5, \ -1.75, -2.0 \rbrace}$.  Next, we reflect these values across axes of symmetry to all other octants of the cube.  
Afterwards, we set $\mathcal{E} = 1$ at the origin.

After multiplying the noise and the spectrum together, we return the result to real space by performing a numerical inverse Fourier transform upon the cube in wavenumber space. This produces an underlying contrast field, $\mathcal{P}_0$, which we then operate upon.  Particularly, we wish to explore the effects on the radiation transfer when the intensity of the clumping produced by the contrast field is varied, so we multiply $\mathcal{P}_0$ by a set of possible amplitude factors, $q = \lbrace{0.0, \ 0.2, \ 0.4, \ 0.6,\  0.8, \ 1.0 \rbrace}$, yielding new fields $q \mathcal{P}_0$.  Then, to roughly approximate a lognormal distribution for the contrasts, we exponentiate the result, transforming the fields to  $\exp{\left(q \mathcal{P}_0\right)}$.  Finally, we apply a normalization factor to create a final contrast field $\mathcal{P}$ where the condition  $\int \rho_{\mathcal{P}}(r) dV = M_{\rm initial} $ is enforced. 

At this point, the completed contrast field $\mathcal{P}$ may be applied to parent envelopes to introduce inhomogeneities in accordance to our spectrum parameter $\beta$ and the amplitude factor $q$.  Note that in the case of $q =0$, $\exp{\left(q \mathcal{P}_0\right)} = 1$ for all points in the envelope , in which case applying $\mathcal{P}$ will simply return the input parent envelope.


\section{Virial Analysis for Internal Motions}\label{virial_analysis}

We wish to supply a brief justification for our definition of the radiation virial term $\cal R$ and its relation to a cloud's centre of mass (CM), so we review here a result from elementary mechanics. Let $\rvec(dm)$ be the position vector of a mass element $dm$ in an inertial reference frame, so that $d\Force = \ddot{\rvec}\,dm$ is the differential of force. Let $\rcm$ be the centre of mass, defined by $M \rcm = \int \rvec\, dm$ (where $M = \int\,dm$ is the cloud mass and all integrals extend over the cloud volume and surface), and let $\rhat= \rvec-\rcm$ be the position relative to the cloud CM, so that $\int \rhat\,dm=0$.   

The trace of the cloud moment-of-inertia tensor is $I = \int \rvec^2\,dm$.  The virial theorem (in Lagrangian form: \citealt{1992ApJ...399..551M}) follows from writing $\ddot{I}/2 = \int \dot\rvec^2 \,dm + \int \rvec \cdot \ddot{\rvec}\,dm$, then recognizing the former term as twice the kinetic energy, and the latter, $\int \rvec\cdot d\Force$, as a series of energies and surface terms.  Given our  definitions, 
\begin{equation}\label{eq:VT_decomposition}
I = M \rcm^2 + \int \rhat^2\,dm \equiv I_{\rm cm} + \hat{I} 
\end{equation}
and therefore $\ddot I = \ddot I_{\rm cm} + \ddot{\hat I}$.  The first of these terms can be written $\ddot I_{\rm cm} = M\dot {\mathbf r}_{\rm cm}^2 + \rcm\cdot \Force$, which is the virial theorem for the motion of the CM.  Defining the internal kinetic energy $\hat{\cal T} = \frac12 \int \dot \rhat^2 \,dM$, the second term is 
\begin{eqnarray}\label{eq:InternalVT}
\ddot{\hat I} &=& \int \dot \rhat^2 \,dM + \int \rhat\cdot(\ddot\rvec-\ddot{\mathbf r}_{\rm cm})\,dm \nonumber \\ 
 &=& 2\hat{\cal T} + \int \rhat\cdot dF + \ddot{\mathbf r}_{\rm cm}  \cdot \int \rhat \,dm 
\end{eqnarray}
in which the final term is zero by virtue of the definition of $\rhat$. 

Equations \ref{eq:VT_decomposition} and \ref{eq:InternalVT} express the fact that the internal motions of the cloud, which represent structural changes, separate cleanly from the CM motion and obey their own virial theorem in which position relative to the CM is the relevant radial vector.   

Like all the other virial terms, the net radiation virial term ${\cal R} = \int \rvec \cdot d\Force_{\rm rad}$ decomposes into two terms: one that appears in the CM virial theorem for $\ddot I_{\rm cm}$, which is ${\cal R}_{\rm cm} = \rcm \cdot F_{\rm rad}$;  and another that appears in the internal virial theorem for $\ddot {\hat I}$, which is 
\begin{equation} \label{eq:Internal_Rad_VirialTerm}
\hat {\cal R} = \int \rhat \cdot d\Force_{\rm rad}.
\end{equation}
As we are primarily concerned with the effect of radiation forces on the internal cloud structure, rather than the magnitude of a radiation-driven rocket effect, we omit the hat and refer to $\hat {\cal R}$ as $\cal R$ in the main text. 

Finally, we note that forces between cloud elements are governed by Newton's third law, so that their contribution to $\Force$ is zero and so is their contribution to $\ddot{I}_{\rm cm}$; for such forces $\ddot{\hat I}$ does not depend on the centre-of-mass location.  Radiation forces are not in this category, because of the reaction force that can accumulate if radiation is incident from one direction, or preferentially escapes in another.  So, for radiation,
the internal virial theorem depends on a cloud's  CM and thus on its mass distribution.  A study like ours must therefore specify the distribution of mass as well as that of opacity, in order to unambiguously define $\hat {\cal R}$. In our calculation of $\rcm$ we make the simplifying assumption that  dust density traces  mass density. 



\appendix

\vskip 1.2in
\bibliographystyle{mnras} 
\bibliography{JumperMatzner2} 


\bsp	
\label{lastpage}
\end{document}